\documentstyle[aps,prl,twocolumn]{revtex}
\begin{document}
\draft
\title{Bloch Oscillation under a Bichromatic Laser: Quasi-Miniband
        Formation, Collapse, and Dynamical Delocalization and Localization}

\author{Ren-Bao Liu and Bang-Fen Zhu}
\address{National Laboratory for Superlattices and Microstructures,
        Institute of Semiconductors,\\  Chinese Academy of Sciences,
        P. O. Box 912, Beijing 100 083, People's Republic of China}
\maketitle

\abstract{A novel DC and AC driving configuration is proposed for
semiconductor superlattices, in which the THz AC driving is provided by
an intense bichromatic cw laser. The two components of the laser,
usually in the visible light range, are near but not exactly
resonant with interband Wannier-Stark transitions, and their frequency
difference equals the Wannier-Stark ladder spacing. Multi-photon processes
with the intermediate states in the conduction (valence) band cause
dynamical delocalization and localization of valence (conduction) electrons,
and the corresponding formation and collapse of the quasi-minibands.}

\normalsize

Semiconductor superlattices (SSL's)\cite{SSL} provide a stage for
low-dimensional physics and for a variety of ultrafast optoelectronic
devices. A typical example for such systems is the well-known
Bloch oscillation (BO) \cite{Zener} and its frequency counterpart,
the Wannier-Stark ladder (WSL) \cite{Wannier}.
Mainly owing to their long artificial periods, SSL's have been the
first systems in which the WSL and BO were observed \cite{WSLBO},
verifying the early predictions \cite{Zener,Wannier} based on the band
theory \cite{Bloch}. Since the BO is accompanied by a time-dependent
dipole momentum, biased SSL's are expected to generate tunable coherent
electromagnetic radiation in the THz range\cite{emission},
which may find extensive applications, e.g., in medical
diagnoses. However, the THz emission from Bloch oscillators has turned
out to be quite weak, and a strong pulsed laser is required for exciting the
wave packets\cite{emission}, which limits its potential applications.

A THz field has been used to drive the BO, as a result the inverse Bloch
oscillators can emit harmonics of the driving
field\cite{DLexperiment,inverse}. As already reported, the biased SSL's
subject to AC-fields exhibit a wealth of interesting physical
effects\cite{AC,perturb0,DCAC,DL,exDL,DDL}. The Hamiltonian of such
systems is time-periodic, thus the eigenstates are the Floquet
states associated with Floquet indices, i.e. quasienergies, forming the
minibands \cite{AC,perturb0,DCAC}. With variation of the THz field
strength, the quasi-miniband width oscillates, and will
collapse at certain field strengths \cite{AC,perturb0,DCAC}.
Correspondingly, the Bloch electrons are
dynamically delocalized and localized\cite{AC,perturb0,DL,exDL,DDL}.
Experiments about these effects, however, have been quite
scarce\cite{DLexperiment,inverse}, because it is not easy to obtain an
intense THz source such as the free electron laser, which has moreover to be
effectively coupled to the SSL's. This also prevents utilization of the THz
driven SSL's in practical devices.

In this Letter, we propose an alternative driving configuration. As
schematically illustrated in Fig. \protect\ref{schematic}, a biased SSL is 
driven by an intense {\em bichromatic cw} laser with its two components at
the frequencies $\omega_1$ and $\omega_2$, and whose average
$\Omega\equiv(\omega_1+\omega_2)/2$ is nearly resonant with $E_g$, the
frequency of the ground-state excitonic Wannier-Stark transition in the
same quantum well, and thus is usually in the {\em visible light} range.
The frequency difference between the two components
$2\omega\equiv|\omega_1-\omega_2|$ is adjusted to $\omega_{BO}$,
the BO frequency or WSL spacing, which is in the THz range. Suppose the two
components have equal weight, then the optical field can be expressed as
\begin{equation}
{\bf E}(t)=2{\bf E}_0\cos(\omega t)\cos(\Omega t). \label{field}
\end{equation}
Here we have assumed the sample to be optically thin and neglected
the spatial dependence of the laser. As will be shown later, this SSL can
be equivalently described by a time-periodic Hamiltonian with the period
$2\pi/\omega$, and in which the dynamical effects are similar to those in
the usual DC-biased and AC driven SSL's\cite{AC,perturb0,DCAC,DL,exDL,DDL}.

At this point, it should be pointed out that although,
atoms excited by two optical fields have been
extensively studied both theoretically and experimentally\cite{DDatom},
and the AC Stark effect in semiconductors driven by an
intense monochromatic laser is also well understood \cite{ACStark},
to the best of our knowledge, there has been no report on Bloch
oscillations driven by a bichromatic laser.

In order to keep the following treatment as simple as possible, we have
made several simplifications. First, we consider only the ground-state exciton
for each Wannier-Stark transition. This is justified because
the laser frequencies can be chosen well below the continuum transitions
in the same quantum well and also by the fact that the ground-state exciton
possesses by far the strongest component of oscillator strength.
Secondly, we neglect the unequal WSL spacing effect\cite{WSLBO,xWSL}
and Fano interference between the continua and
the embedded discrete states\cite{FR} induced by the Coulomb interaction.
Usually, the excitonic coupling affects significantly the dynamical
delocalization or localization\cite{exDL,DDL}, this however will be
taken into account in our later work. Thirdly, the Zener tunneling and
band mixing effects are not included. And lastly, the laser is
assumed detuned for transitions other than those between the
highest valence and the lowest conduction minibands. All these assumptions
will in some way affect our results. Even so, the
principle as well as main conclusions will not be qualitatively modified.

With these simplifications, the Hamiltonian can be mapped into a
one-dimensional model with two-band as well as nearest-neighbor
coupling. With a well-justified rotating wave approximation for
the optical coupling and the unitary transformation
$$S=\exp[i\Omega t\sum_n
(a^{\dag}_{{\rm c}n}a_{{\rm c}n}-a^{\dag}_{{\rm v}n}a_{{\rm v}n})/2],$$
the Hamiltanian in the rotating frame is
\begin{eqnarray}
H=\sum_{n} &&(n\omega_{BO}+\frac{\varepsilon_0}{2})
 a^{\dag}_{{\rm c}n}a_{{\rm c}n}
-\frac{\Delta_{\rm c}}{4}a^{\dag}_{{\rm c}n}(a_{{\rm c}n+1}+a_{{\rm c}n-1})
\nonumber\\
+&&(n\omega_{BO}-\frac{\varepsilon_0}{2})a^{\dag}_{{\rm v}n}a_{{\rm v}n}
+\frac{\Delta_{\rm v}}{4}a^{\dag}_{{\rm v}n}(a_{{\rm v}n+1}+a_{{\rm v}n-1})
\nonumber\\
-&&\chi\cos(\omega t)(a^{\dag}_{{\rm c}n}a_{{\rm v}n}+ {\rm H.c.}),
\label{H}\end{eqnarray}
where $a^{\dag}_{{\rm c}n}$ ($a^{\dag}_{{\rm v}n}$) stands for the
operator creating a conduction (valence) electron at the $n$th
site, $\Delta_{\rm c}$ ($\Delta_{\rm v}$) is the conduction (valence)
miniband width, $\varepsilon_0\equiv E_g-\Omega$ is the detuning, and
$\chi\equiv{d E_0}$ is the interband optical transition strength ($d$ is the
interband dipole matrix element along the ${\bf E}_0$ direction).
Compared to the usual THz driving WSL system,  the driving
force in the present configuration is essentially an interband
rather than an intraband process.

The time-periodic system described by Eq. (\protect\ref{H})
has no stationary solution, and the eigenstates are time-periodic
Floquet states $|q,m\rangle\equiv\exp(im\omega t)|q\rangle$
with period $T=2\pi/\omega$, which satisfy
$$(H-i\partial_t)|q,m\rangle=\varepsilon_{q,m}|q,m\rangle,$$
where $\varepsilon_{q,m}\equiv\varepsilon_q+m\omega$
is the Floquet index, i.e. quasienergy.
The Floquet states and indices can be obtained by diagonalizing the
secular equation $$U(T,0)|q\rangle=\exp(-i\varepsilon_q T)|q\rangle,$$
where the propagator $$U(T,0)\equiv \hat{T}\exp[-i\int^T_0H(t)dt].$$

The Hamiltonian (\protect\ref{H}) can be decomposed into a sum of
mutually commutative $2\times2$ matrices in the accelerating
quasimomentum representation $\{|\alpha,\tilde{k}\rangle|\alpha=$c, v\}
with the transformation
$|\alpha,\tilde{k}\rangle=|\alpha,k-\omega_{BO}t\rangle$
(hereafter the superlattice period is assumed to be unity). That is
$H=\sum_k H_k$, with
\begin{equation}
H_k=\sum_{{\rm \alpha= c, v}}
\varepsilon_{{\rm \alpha}\tilde{k}}a^{\dag}_{{\rm \alpha}\tilde{k}}
a_{{\rm \alpha}\tilde{k}}
-\chi\cos(\omega t)(a^{\dag}_{{\rm c}\tilde{k}}a_{{\rm v}\tilde{k}}
+{\rm H.c.}),
\end{equation}
where $\varepsilon_{{\rm c}\tilde{k}}=
(\varepsilon_0-\Delta_{\rm c}\cos\tilde{k})/2$
and $\varepsilon_{{\rm v}\tilde{k}}=
(-\varepsilon_0+\Delta_{\rm v}\cos\tilde{k})/2$
are the conduction and valence miniband dispersions, respectively.
Thus each quasimomentum $k$ corresponds to two sets of Floquet states
$|\pm,m\rangle_k$ with the quasienergies $\varepsilon_{k\pm,m}$.

Before we present our numerical results, the quasi-minibands are
analytically evaluated in the weak and strong optical coupling limits,
so as to give some physical insight. From now on, we will set $E_g=\Omega$.

When $\chi=0$, i.e. there is no optical field, $H_k$ is already diagonalized,
and the quasienergies produce the well-known WSL \cite{Wannier}.

When $\chi$ is small, the optical coupling $$H_{1,k}\equiv
-\chi\cos(\omega t)(a^{\dag}_{{\rm c}\tilde{k}}a_{{\rm v}\tilde{k}}
+{\rm H.c.})$$ can be treated as a perturbation to
$$H_{0,k}\equiv
\varepsilon_{{\rm c}\tilde{k}}a^{\dag}_{{\rm c}\tilde{k}}
a_{{\rm c}\tilde{k}}+
\varepsilon_{{\rm v}\tilde{k}}a^{\dag}_{{\rm v}\tilde{k}}
a_{{\rm v}\tilde{k}}.$$
Similar to the steady case, a perturbation theory can be developed,  
in which  the matrix element for time-periodic functions is defined as 
$$\langle\langle q,m|H|q',m'\rangle\rangle\equiv\frac{1}{T}\int^T_0
\langle q,m|H|q',m'\rangle dt$$ in the Hilbert subspace\cite{perturb0}.
In terms of unperturbed Floquet states
$$|\pm\rangle_k=\exp[\pm i\Delta_{\rm c/v}/(2\omega_{BO})
\sin(\omega_{BO}t-k)]|{\rm c/v},\tilde{k}\rangle$$
 with the quasienergies $\varepsilon_{k\pm}^{(0)}=0$,
the optical coupling is expressed as
\begin{eqnarray}
&& _k\langle\langle+,m|H_{1,k}|+,m'\rangle\rangle_k=
_k\langle\langle-,m|H_{1,k}|-,m'\rangle\rangle_k=0,\label{matrix0}\\
&& _k\langle\langle-,m|H_{1,k}|+,m'\rangle\rangle_k=
_k\langle\langle+,m'|H_{1,k}|-,m\rangle\rangle_k^\ast\nonumber\\
&&=-\frac{\chi}{2}\sum_{n,\pm}\delta_{n,{\frac{m-m'\pm1}{2}}}
J_n(\frac{\Delta}{2\omega_{BO}})e^{-ink},\label{matrix}
\end{eqnarray}
where $\Delta\equiv\Delta_{\rm c}+\Delta_{\rm v}$
is the combined miniband width,
$n$ is an integer number, and $J_n(x)$ is the $n$th order Bessel
function of the first kind.
According to Eq. (\protect\ref{matrix0}) and (\protect\ref{matrix}), the
quasienergy spectrum depends on only $\Delta$ rather than the ratio
$\Delta_{\rm c}/\Delta_{\rm v}$.

From the symmetry of the matrix elements,
the quasienergies $\varepsilon_{k\pm}$ satisfy the relation
$\varepsilon_{k\pm}=-\varepsilon_{k\mp}=\varepsilon_{\pi-k\mp}=
\varepsilon_{k-\pi\mp}=\varepsilon_{-k\pm}$, so each Floquet state
is 4-fold degenerate, and only the quasienergies for $k\in[0,\pi/2]$
need to be calculated to obtain the full spectrum information.

Since the first-order modification for the quasienergies vanishes
(Eq. (\protect\ref{matrix0})), the second-order perturbation gives
\begin{equation}
\varepsilon_{k\pm}=\mp\chi^2\cos k
\sum_{n=1}^{+\infty}\frac{J_n(\frac{\Delta}{2\omega_{BO}})
J_{n-1}(\frac{\Delta}{2\omega_{BO}})}{(2n-1)\omega},\label{second}
\end{equation}
which forms a miniband. It can be recognized at once that, except for a
prefactor, the quasi-miniband has a dispersion relation identical to that
of the original minibands. The discrete WSL is broadened, indicating
dynamical delocalization of the originally localized Wannier-Stark states.

An intuitive picture favors our understanding of the dynamical delocalization.
As depicted in Fig. \protect\ref{schematic} (b), a valence band electron
in a Wannier-Stark state can hop to its neighbor states by first absorbing
a photon from one component of the bichromatic laser and then emitting a
photon to the other component, though both components are
off-resonant with the interband transitions.
The square law dependence of the quasi-miniband width on the optical
coupling strength $\chi$ results from this second-order process, in
which a conduction band electron plays the role of an intermediate state.
Clearly, the Bessel functions $J_n(\frac{\Delta}{2\omega_{BO}})$ and
$J_{n-1}(\frac{\Delta}{2\omega_{BO}})$ in Eq. (\protect\ref{second})
reflect the two virtual transitions between the intermediate state
and the initial and final Wannier-Stark states.
A conduction electron can similarly be delocalized.

With increasing laser intensity, multi-photon processes become
more and more important, and the perturbation treatment for the
optical coupling will fail eventually. Thus, when $\chi\gg\Delta$,
the $H_{0,k}$ should be regarded as the perturbation to
$H_{1,k}$  \cite{perturb0}. Since eigenstates of $H_{1,k}$ are
$$|\pm\rangle_k=\exp[\pm i{\chi}/{\omega}\sin(\omega t)]\sqrt{2}/2
(|{\rm c},\tilde{k}\rangle\pm|{\rm v},\tilde{k}\rangle),$$ 
the first-order degenerate perturbation gives the quasi-miniband dispersion as
\begin{equation}
\varepsilon_{k\pm}=\pm\frac{\Delta}{4}J_2(\frac{2\chi}{\omega})
\cos k .\label{first}\end{equation}        
In this strong coupling limit, the quasi-miniband width oscillates according
to a second order Bessel function with the argument of $2\chi$ over $\omega$,
and the quasi-miniband collapses at such laser intensities that
$2\chi/\omega$ is a root of the Bessel function, which is very similar
to the case of intra-miniband driving by a THz field.
The Taylor expansion of the Bessel function in Eq. (\protect\ref{first})
contains only even powers of $\chi$, indicating that each virtual photon
absorption is necessarily followed by a virtual
photon emission, and no process with an odd number of photons is involved.
This again confirms our intuitive picture invoked to understand the dynamical 
delocalization.

The formation and collapse of the quasi-miniband
can be clearly seen from Fig. \protect\ref{spectrum}, in which the
quasienergies are numerically calculated as functions of the optical coupling 
strength for 21 $k$ points taken evenly from 0 to $\pi/2$.   
For comparison, the perturbation results for $k=0$ are also plotted.
In the weak optical coupling limit ($2\chi/\omega<2$), the second order
perturbation of $H_{1,k}$  (dashed lines) gives results well consistent
with the numerical ones. In the strong coupling limit, the first order
perturbation of $H_{0,k}$ is also a quite good approximation, especially
for smaller $\Delta$, where the requirement for the perturbation treatment
$\chi\gg\Delta$ is better met. 

Comparing the numerical results with the perturbation calculations, two
points need mentioning. First, the quasi-miniband collapses at
$2\chi/\omega\approx$4.16, 7.84, etc. for $\Delta=33$ meV, and
at $2\chi/\omega\approx$4.96, 8.32, etc. for $\Delta=16.5$ meV,
which correspond to laser intensities a little weaker than the
roots of the Bessel function. We think it comes from the underestimated
localization effect of the DC field by treating $H_{0,k}$ as a
perturbation. Secondly, the numerically calculated band collapse is not as
complete as that in perturbation calculations, because of the interaction
between different sidebands of the Floquet states of $H_{1,k}$.

To show the dynamical delocalization and localization of a bichromatic
laser driven electron, we have calculated the evolution of an initially
localized electron by integrating the Shr\"{o}dinger equation. The wave packet
width $W$ defined to characterize the localization of an electron is
$$W^2\equiv\sum_n(n-r_0)^2(\langle a^{\dag}_{{\rm c}n}a_{{\rm c}n}\rangle
+\langle a^{\dag}_{{\rm v}n}a_{{\rm v}n}\rangle),$$  where
$$r_0\equiv\sum_nn(\langle a^{\dag}_{{\rm c}n}a_{{\rm c}n}\rangle
+\langle a^{\dag}_{{\rm v}n}a_{{\rm v}n}\rangle).$$
When $t=0$, the electron is put at the valence site $n=0$ \cite{note}. With
such an initial condition, the wave packet center $r_0$ is found to
be always zero for any laser intensity.

In Fig. \protect\ref{localization}, $W$ is shown as a function of time within
the first ten periods for three laser intensities. Without the optical
field (Fig. \protect\ref{localization} (a)), the wave packet
width $W$ oscillates sinusoidally with time with a static center-of-mass,
which is just the breathing mode of the BO\cite{breathing} and the
well-known Wannier-Stark localization. The two non-zero laser intensities
correspond to approximately maximum quasi-miniband width and band collapse,
respectively. When the quasi-miniband width is a maximum
($2\chi/\omega=3.04$, Fig. \protect\ref{localization} (b)), $W$ increases
linearly with time with a slight oscillation superimposed, which clearly
evidences the dynamical delocalization of the initially localized electron.
When the quasi-miniband collapses ($2\chi/\omega=4.96$,
Fig. \protect\ref{localization} (c)), $W$ oscillates with time
significantly and quasi-periodically, whose minima
are slightly shifted upward due to imperfect band collapse. It is
obvious that the dynamical localization is realized.

Note that the dynamical localization length is larger than the
Wannier-Stark localization length by a factor of  about
$\Delta/\Delta_{\rm v}$, which suggests  a similar band-width dependence
for both kinds of localization. This behavior is also exhibited for dynamical
delocalization, which, as shown in Fig. \protect\ref{localization} (b),
compares very well to the diffusion of a free Bloch electron provided
identical bandwidths ($\approx$3.9 meV) are adopted.

The population of the conduction miniband
$f_{\rm c}\equiv\sum_n\langle a^{\dag}_{{\rm c}n}a_{{\rm c}n}\rangle$
is also plotted for $2\chi/\omega=3.04$ (the dotted line). $f_{\rm c}$
oscillates nearly periodically with time due to virtual excitation.
When a valence band electron is excited into the conduction states,
it can hop to neighbor sites, thus $W$ ascends by a step, which confirms
once more our intuitive picture for the dynamical delocalization process.

In summary, we have predicted the quasi-miniband formation and collapse,
and correspondingly the dynamical delocalization and localization for
an electron in a SSL driven by DC and AC fields in a novel configuration,
in which the THz AC driving is provided by an intense bichromatic visible
laser with the frequency difference of the two components equal to the WSL
spacing. Our findings suggest that the experiments on the dynamical effects
can be carried out without expensive THz sources like free
electron lasers. Suppose $\omega_{BO}=2\omega=10$ meV, an optical coupling
of 12.4 meV is enough to realize the dynamical localization, which should
be accessible already in many laboratories. It should be noted that the
two components of the bichromatic laser can be replaced by two independent
lasers with small phase fluctuation.
The possible THz emission from so driven Bloch oscillators, requiring
neither an intense THz source nor pulse optical excitation, may find
broad applications. In addition, this interband driving configuration
introduces  several interesting physical problems such as the excitonic
effect and excitation-induced many body correlation,
which are absent in the usual intraminiband THz driving configuration.

The authors are grateful to Professor Kun Huang and A. Rhys for their
critical reading of the manuscript. This work was supported by the
National Science Foundation of China and the QiuShi Science \& Technology
Foundation of Hong Kong.

\begin{figure}
\caption{Schematics for (a) the DC- and bichromatic laser driven SSL system, and
 (b) the two components of the bichromatic laser (solid arrows), the
 WSL, and the virtual transitions leading to dynamical delocalization
 (dotted arrows).}
\label{schematic}
\end{figure}  

\begin{figure}
\caption{Quasienergies as functions of the optical coupling strength for
 21 equally spaced $k$'s from 0 to $\pi/2$ with the following parameters:
 $\omega_{BO}=2\omega=10$ meV, and (a)$\Delta_{\rm c}=10\Delta_{\rm v}=30$
 meV, (b)$\Delta_{\rm c}=10\Delta_{\rm v}=15$ meV. The
 perturbation results for $k=0$ at the weak and strong optical coupling
 limits are plotted in dashed and dotted lines, respectively.}
\label{spectrum}
\end{figure}  

\begin{figure}
\caption{Wave packet width (solid lines)
 as functions of time within the first ten periods
 for $2\chi/\omega=0$, 3.04, and 4.96 in (a), (b), and (c), respectively.
 Other parameters are the same as in Fig. \protect\ref{spectrum} (b). In
 (b), the dashed line stands for wave packet width of a free Bloch
 electron which is initially localized, and the dotted line is the
 conduction electron population.}
\label{localization}
\end{figure}

\begin{references}
\bibitem{note} We have tried other initial conditions such as
 the Gaussian distribution and the Wannier-Stark state, the main effects
 are unchanged.
\end{references}
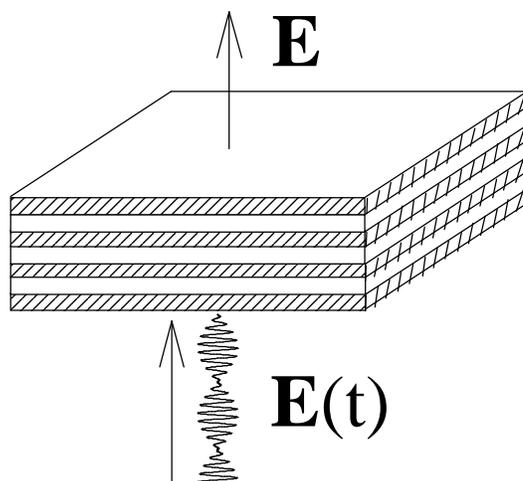
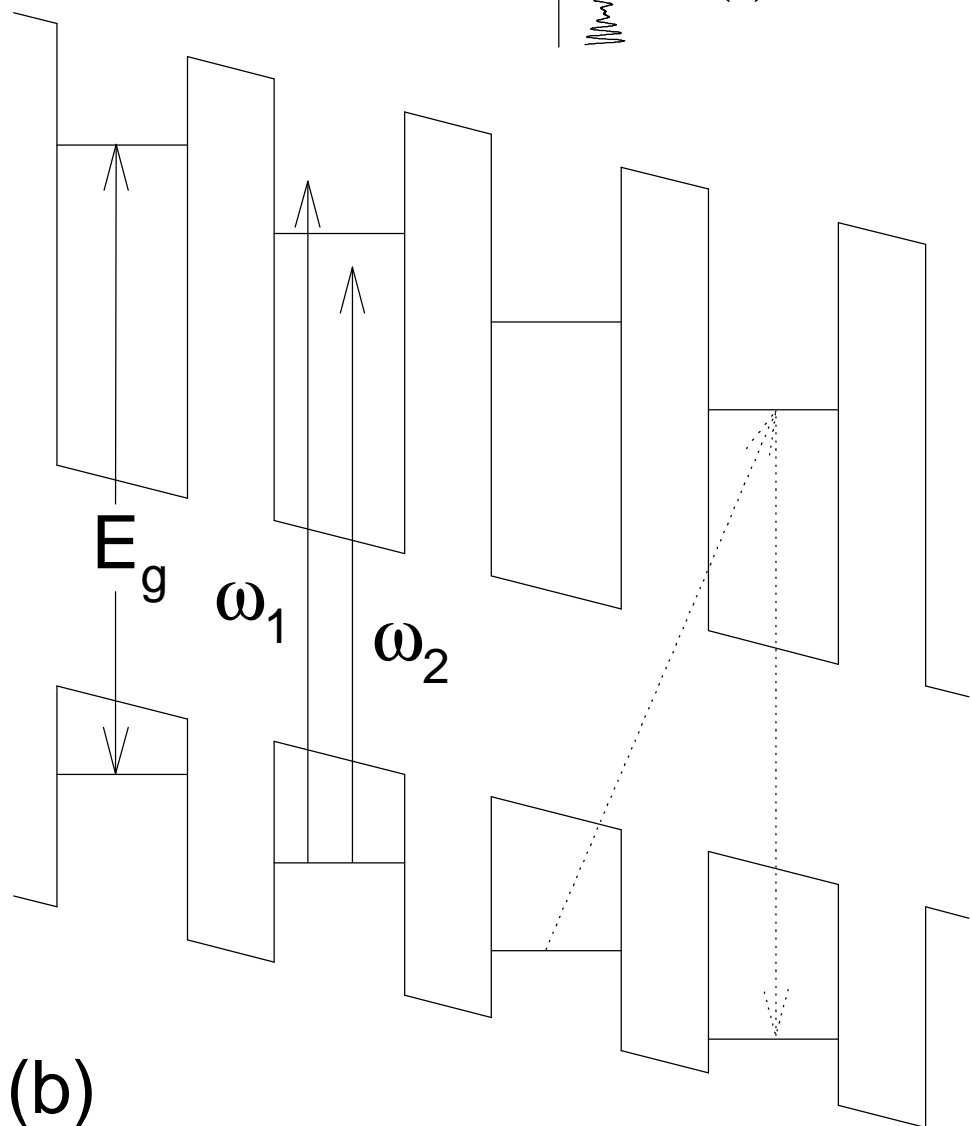
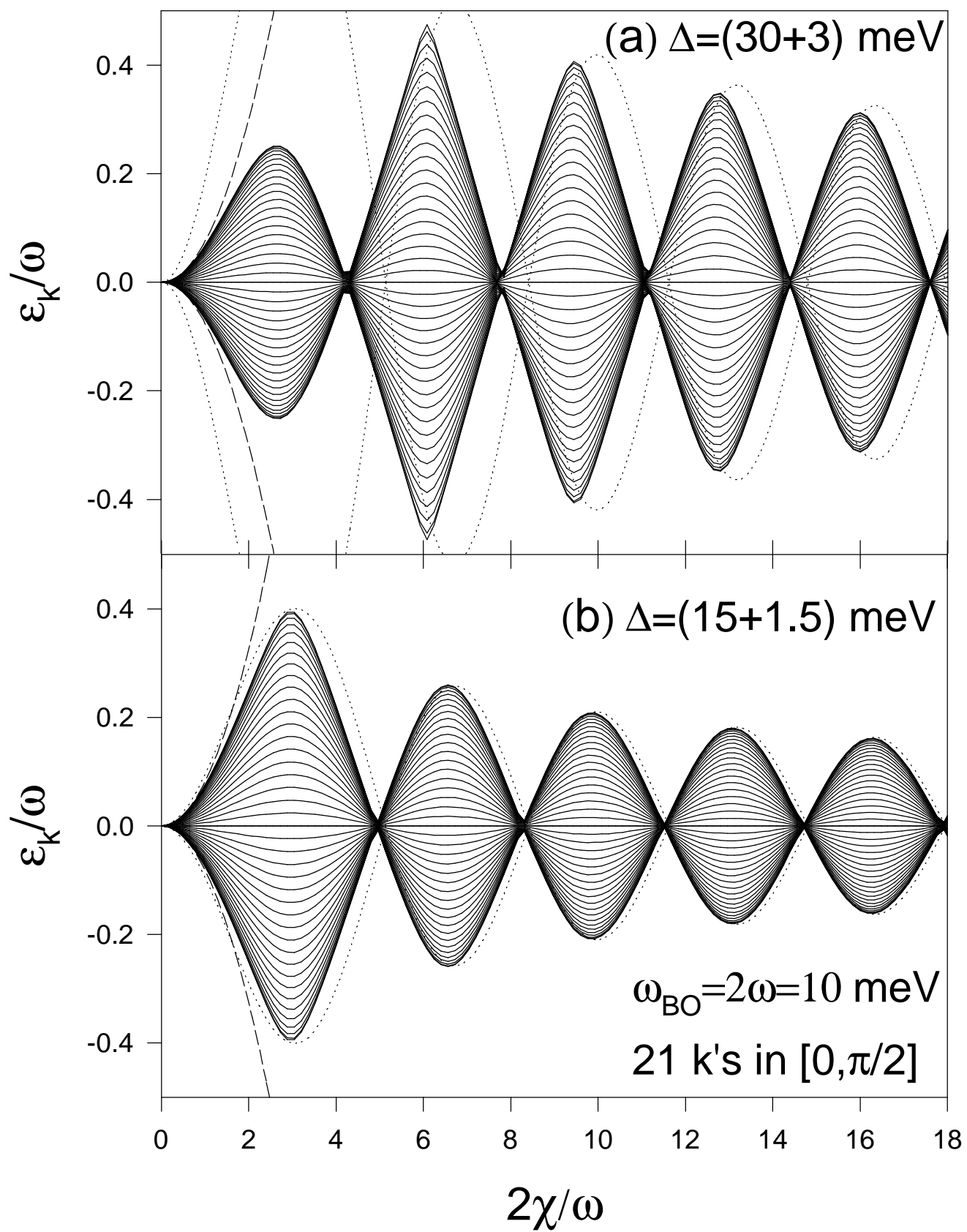
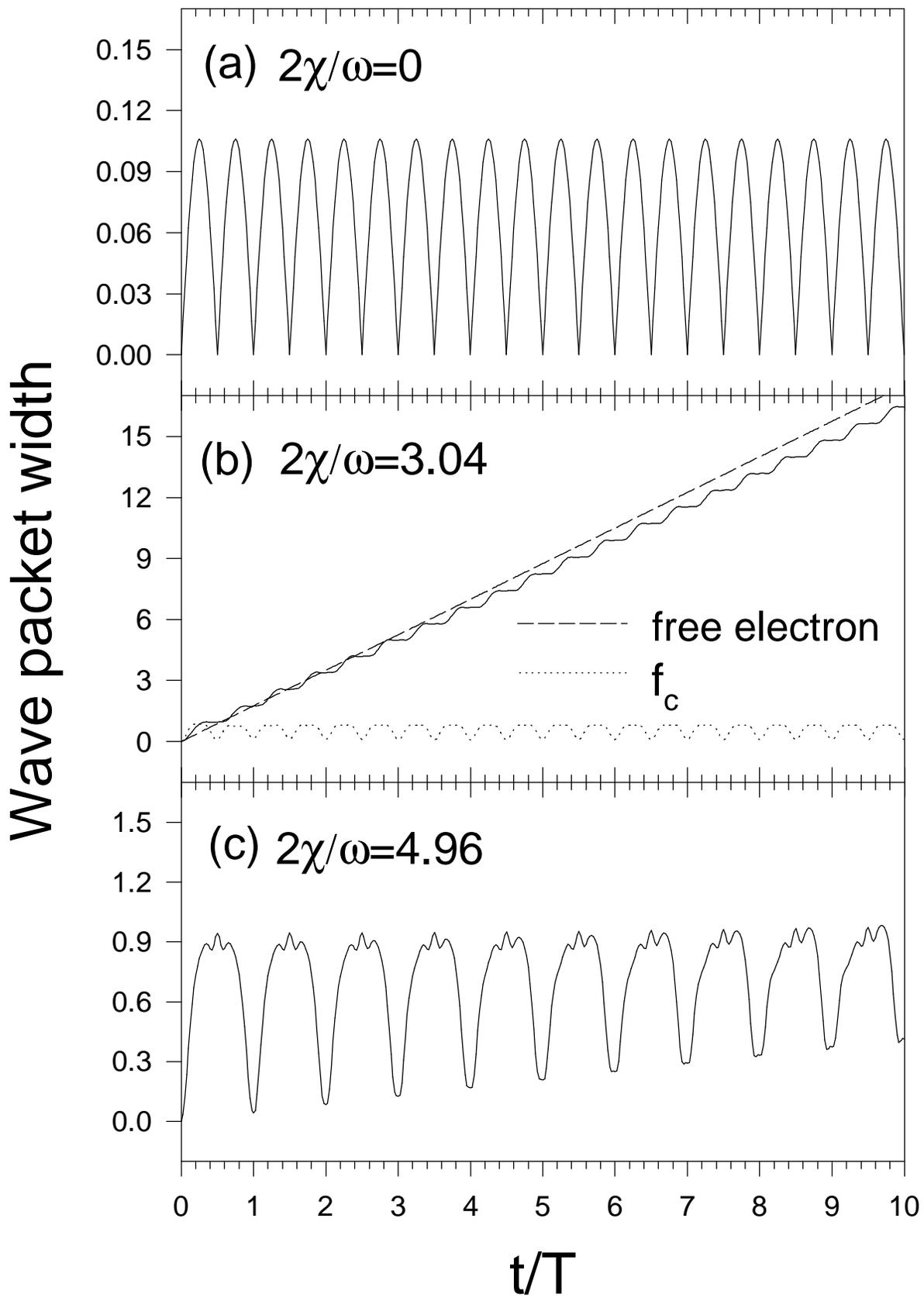
\end{document}